\documentclass[conference]{IEEEtran}
\IEEEoverridecommandlockouts
\usepackage{cite}
\usepackage{amsmath,amssymb,amsfonts}
\usepackage{algorithmic}
\usepackage{multirow}
\usepackage{graphicx}
\usepackage{textcomp}
\usepackage{xcolor}
\usepackage{diagbox}
\def\BibTeX{{\rm B\kern-.05em{\sc i\kern-.025em b}\kern-.08em
    T\kern-.1667em\lower.7ex\hbox{E}\kern-.125emX}}
\usepackage{listings}
\usepackage{comment}
\usepackage{subfig}
\usepackage{todonotes}
\usepackage{hyperref}
\usepackage{array}
\usepackage{siunitx}
\usepackage{mathrsfs}
\usepackage{fancyvrb}

\usepackage[ruled,vlined]{algorithm2e}
\newcolumntype{L}[1]{>{\raggedright\arraybackslash}p{#1}}
\newcolumntype{C}[1]{>{\centering\arraybackslash}p{#1}}
\newcolumntype{R}[1]{>{\raggedleft\arraybackslash}p{#1}}

\newcommand{\red}[1]{\textcolor{red}{#1}}

\begin{document}

\title{DeepFreeze: Cold Boot Attacks and High Fidelity Model Recovery on Commercial EdgeML Device}

\makeatletter
\newcommand{\linebreakand}{%
  \end{@IEEEauthorhalign}
  \hfill\mbox{}\par
  \mbox{}\hfill\begin{@IEEEauthorhalign}
}
\makeatother

\ifdefined\ANONYMOUS
\author{

}
\else
\author{
\IEEEauthorblockN{Yoo-Seung Won\textsuperscript{1}, Soham Chatterjee\textsuperscript{2}, Dirmanto Jap\textsuperscript{1}, Arindam Basu\textsuperscript{2,3}, Shivam Bhasin\textsuperscript{1}}
\IEEEauthorblockA{\textsuperscript{1}\textit{Temasek Laboratories},
\textit{Nanyang Technological University},
Singapore. \\
\textsuperscript{2}\textit{School of Electrical and Electronic Engineering}, 
\textit{Nanyang Technological University}, 
Singapore. \\
\textsuperscript{3}\textit{Department of Electrical Engineering}, 
\textit{City University of Hong Kong}
Hong Kong. \\
Email: \{yooseung.won,djap,sbhasin\}@ntu.edu.sg, soham004@e.ntu.edu.sg, arinbasu@cityu.edu.hk
}
}
\fi

\maketitle

\begin{abstract}

EdgeML accelerators like Intel Neural Compute Stick 2 (NCS) can enable efficient edge-based inference with complex pre-trained models. The models are loaded in the host (like Raspberry Pi) and then transferred to NCS for inference. In this paper, we demonstrate practical and low-cost cold boot based model recovery attacks on NCS to recover the model architecture and  weights, loaded from the Raspberry Pi. The architecture is recovered with 100\% success and weights with an error rate of 0.04\%. The recovered model reports maximum accuracy loss of 0.5\% as compared to original model and allows high fidelity transfer of adversarial examples. We further extend our study to other cold boot attack setups reported in the literature with higher error rates leading to accuracy loss as high as 70\%. We then propose a methodology based on knowledge distillation to correct the erroneous weights in recovered model, even without access to original training data. The proposed attack remains unaffected by the model encryption features of the OpenVINO and NCS framework.

\end{abstract}


\section{Introduction}
\label{sec:intro}

The tremendous success of Deep learning (DL) across varied fields such as image recognition to natural language processing has fuelled a new wave of artificial intelligence (AI). Coupled with the exponential increase in the number of sensor nodes in the internet of things (IoT), this has brought about new requirements for AI hardware where computing is done at the network edge close to the sensor \cite{Edge-review1}. Dubbed as EdgeML, it is a growing research area that aims to deploy models on computationally weak devices while still maintaining model accuracy and performance requirements. Deploying models on microcontrollers and other edge accelerators like Intel Neural Compute Stick 2 (NCS)~\cite{ncs2}, ensures reduced latency and data transmission requirements.

However, edge devices are exposed to many security concerns due to the attacker having physical access to the device \cite{DBLP:conf/uss/BatinaBJP19}. The three major concerns relate to data privacy or safety for end users, fault injection to induce misclassification and IP theft for AI service providers. A lot of works have focused on input manipulation to create `adversarial' inputs that will be misclassified with high probability \cite{adversarial} while others have focused on injecting physical faults \cite{DAC2020_CCH}. In this work, we concern ourselves with the latter problem of IP theft.  Due to the usage of proprietary training data and expertise of machine learning scientists in developing DNN models, these are valuable IP that attackers are interested in acquiring\cite{DAC2020_reverse}. Moreover, this knowledge can be leveraged by the attacker to launch adversarial attacks as well.

\begin{figure*}[ht]
  \centerline{\includegraphics[width=0.8\linewidth]{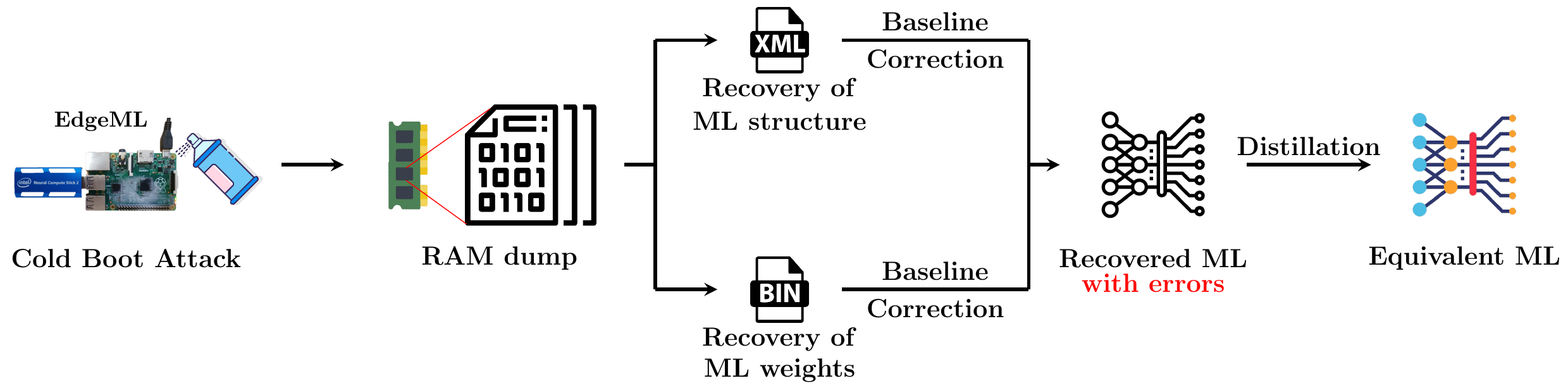}}
  \caption{Attack Flow for Model Recovery on Commercial EdgeML Device.}
  \label{fig:attack_flow}
\end{figure*}

Earlier works have demonstrated attacks exploiting physical access, enabled by EdgeML, such as side-channel~\cite{DBLP:conf/uss/BatinaBJP19}, faults~\cite{DAC2020_CCH} \textit{etc}. However, these attacks have associated cost owing to the required equipment. In this work, we demonstrate practical and extremely low-cost method for model recovery through cold boot attack on EdgeML devices. Cold boot attacks~\cite{halderman2009lest} steal sensitive information like passwords, encryption keys in SRAM by forcing memory to extreme low temperatures. At such temperatures, the data persists in SRAM even after power off for several minutes allowing malicious recovery. 

EdgeML device when executing model inference stores the sensitive trained model in SRAM which becomes a target of cold boot attack. The built in model encryption in NCS tool flow does not protect this attack as the model must be decrypted before execution\footnote{The vulnerability was responsibly disclosed to Intel in March 2021. Intel recommends use of secure hardware enclaves to mitigate such attack. Availability of secure hardware enclaves is not always guaranteed on low-cost host setup and must be carefully evaluated based on the sensitivity of the application.}. The whole attack process is illustrated in Fig.~\ref{fig:attack_flow}. The model recovered with cold boot attack has certain errors. We then propose a methodology based on knowledge distillation to correct the model errors in order to achieve comparable accuracy.

\subsection{Our Contributions}

The main contributions of this work are as follows:
\begin{enumerate}
    \item We propose first cold boot attack on EdgeML paradigm enabling secret model recovery. The attack is low cost and demonstrated on a EdgeML setup composed on NCS as an AI accelerator and Raspberry Pi as the host, where host becomes the primary target. The proposed attack also bypasses the built in model encryption in NCS tool flow. 
    \item With the achieved low error rate on our attack setup, we report a 100\% success in recovery of model architecture and weights are recovered with a minor error of 0.04\%. The recovered model are comparable to original model and results in a relative accuracy drop of under $0.005$ on $6$ different models.
    \item We achieve high-fidelity extraction of the model, according to the taxonomy of~\cite{jagielski2020high}, which provides advanced capability to an adversary like transferring adversarial examples with a high success rate.
    \item As the errors in cold boot attack can increase drastically with a minor disturbance in experimental environment, we report model recovery at higher error rates as reported in previous works~\cite{halderman2009lest} showing relative accuracy drop as high as $0.7$. We thus propose a novel model correction approach based on dropout knowledge distillation bringing relative accuracy drop to under 10\%. The main advantage of the proposed approach is that it works even without access to original training dataset.  
\end{enumerate}

The codes for the experiments reported in this work is made publicly available at \href{https://github.com/soham96/DeepFreeze}{https://github.com/soham96/DeepFreeze}.

\section{Preliminaries}
\label{sec:prelim}

In this section, we recall the background concepts that will be used for the rest of the paper.

\subsection{Cold Boot Attack}
The cold boot attacks~\cite{halderman2009lest} exploit the data remanence property of static random access memory (SRAM) memory cell to recover sensitive data.
SRAM which is considered volatile, loses data instantly on power-off.
However, if the memory is forced into extreme cold temperatures, the data on the memory will decay at a much lower rate, allowing enough time for an adversary to recover it. The SRAM can contain sensitive information like passwords, PIN, login credentials, encryption keys \textit{etc}. Unauthorised access to SRAM data through cold boot attacks can lead to serious security breach in the target system. The practicality of cold boot attacks have been widely demonstrated on platforms like laptop~\cite{halderman2009lest}, desktop~\cite{olle2018anice}, scrambled RAM~\cite{bauer2016lest}, smartphones~\cite{tilo2014frost} and IoT device~\cite{won2020practical}. 
Depending on the temperature of the memory, the data stored in memory can potentially retain even though losing the power. Hence, an adversary can recover the data using these properties if the sensitive/private data are stored in the memory. In terms of adversary assumption, we can divide into two categories, depending on the RAM separation from the main board.

If the RAM can be separated from the main board (\textit{e.g.} desktop and laptop), the adversary can remove the victim memory and connect to another machine for recovering the data. If the memory cannot be separated due physical constraints (like stacked memory in chip package), a vulnerable boot sequence can be exploited to recover the sensitive data from cold boot~\cite{won2020practical}. In this paper, we concentrate on the vulnerability of Raspberry Pi against cold boot attack in order to investigate the ML model recovery.

\subsection{Model Extraction Attack}

In machine learning, training a model requires lots of resources with associated cost. As a result, a well trained model has commercial importance. This has led to the rise of Machine Learning as a Service (MLaaS) where big companies like Amazon, Google, \textit{etc} have trained efficient models owing to their access to huge amount of training data. The models are available on a pay per use basis. A theft/recovery of such trained models by an adversary incurs direct losses to the model owner.
 
Model recovery or model extraction is an active topic in the area of machine learning.
Tramer \textit{et al.}~\cite{tramer2016stealing} demonstrated that it is possible to duplicate the functionality of a  black box model API, without any prior knowledge of model parameters and training data. The adversary observes responses to known queries and aims to estimate a model which is functionally close or equivalent to the black box model. Later, with the adoption of ML on edge devices (EdgeML), the use case enabled physical access to the target inference device. Physical access allowed new attack vectors like side-channels~\cite{DBLP:conf/uss/BatinaBJP19} and faults~\cite{breier2020sniff}. Some of these attack exploit: power~\cite{DBLP:journals/corr/abs-1910-13063}, electromagnetic~\cite{DBLP:conf/uss/BatinaBJP19}, and timing~\cite{DBLP:conf/dac/HuaZS18} leakages to recover model parameters. Faults attacks have also been shown to enable model recovery~\cite{breier2020sniff}.
To the best of our knowledge, using cold boot attack for model recovery has not been investigated before.

The idea is to recover the trained model.  The first model extraction attack was proposed by Tramer \textit{et al.} \cite{tramer2016stealing}. In this case,  they assume black box model, with no prior knowledge of the model parameter and training data. The aim is to duplicate the functionality of the target model, by observing the output of the model. In \cite{jagielski2020high}, the author proposed a taxonomy of the model extraction attacks on machine learning. Basically, the attacks can be categorized as follow (with addition of exact extraction in this work):
\begin{itemize}
    \item \textit{Exact Extraction:} when the extracted model have same architecture and weights as the original,
    \item \textit{Functionally Equivalent Extraction:} a slightly weaker assumption, where the output of both models only have to agree to be the same for all the elements from the domain,    
    \item \textit{Fidelity Extraction:} where the extracted model be the one that maximise the similarity function with the original model for a target distribution. The functionally equivalent extraction is special case, where it  achieves a fidelity of 1 on all distributions and all distance functions, and
    \item \textit{Task Accuracy Extraction:} where the extracted model only has to match (or exceed) the accuracy of the target model.
\end{itemize}

While task accuracy extraction is the most commonly seen goal, exact extraction is observed in attacks leveraging physical means~\cite{DBLP:conf/uss/BatinaBJP19,breier2020sniff} at the cost of high-end equipment. High fidelity extraction was exploited in~\cite{jagielski2020high} to validate transfer of adversarial examples and is considered a stronger extraction as compared to task accuracy.

\subsection{Knowledge Distillation}
Knowledge Distillation (KD) was first introduced by Hinton \textit{et al.}~\cite{hinton2015distilling} as a model compression technique where the authors tried to distill the knowledge learned by a larger teacher network into a smaller student network. The student learns using the output class probabilities or the soft-outputs from the teacher network. Hinton \textit{et al.}, also proposed the use of a softmax temperature factor and later works have proposed using intermediate layer outputs to increase the amount of information available to the student for training \cite{gou2021knowledge}. KD has also been combined with other techniques like pruning and quantization to get further compression. KD has been used as a method to increase the accuracy of the student network by using a trained teacher model to generate pseudo labels for a large unlabelled image dataset \cite{gou2021knowledge}. This has been shown to improve the accuracy on ImageNet by 2\%. 

\subsection{Related Works}

Several model recovery attacks have been proposed in literature~\cite{jagielski2020high,tramer2016stealing}. 
With the adoption of EdgeML, attacks targeting the physical layer have emerged leveraging on physical access through side-channel analysis, fault injection \textit{etc}. The approach based on side-channel attacks exploit physical leakages, such as power and electromagnetic emanation (EM)~\cite{DBLP:conf/uss/BatinaBJP19,DBLP:journals/corr/abs-1903-03916,yu2020deepem} to recover parameters like architecture, weights of the target model\footnote{Cache based and other remote side-channel attacks are not considered in light of EdgeML devices.}. Other attacks exploiting physical access include faults~\cite{breier2020sniff}, where the weights of last layers were retrieved by precise fault injection. As per our knowledge, using cold boot attack for model recovery has not been investigated yet. In Tab.~\ref{related} we compare the common techniques leveraging physical access for model recovery. The cost of the attack is estimated based on equipment reported in the respective paper. The extraction methods refers to taxonomy of~\cite{jagielski2020high} as previously described, with the addition of exact extraction, which to our knowledge is specific to methods leveraging physical access.

\begin{table}[t]
    \centering
    \fontsize{6.5}{7.5}\selectfont
    \caption{Comparison with Related Works Leveraging Physical Access}
    \label{related}
    \begin{tabular}{c|c|c|c}
    \hline
        Technique & Recovered Parameters & Extraction Method & Cost(\$)  \\\hline
         EM Side-Channel~\cite{DBLP:conf/uss/BatinaBJP19} & Architecture + Weights & Exact & 10k-100k\\
         Power Side-Channel~\cite{dubey2020maskednet} & Weights & Exact & 1k-10k\\
         EM Side-Channel~\cite{yu2020deepem} & Architecture + Weights & Task Accurate & 10k-100k\\
         Faults~\cite{breier2020sniff} & Weights (last layer) & Exact & $>$100k\\
         Cold boot [This Work] & Architecture + Weights & High Fidelity & $<$10\\\hline
    \end{tabular}
\end{table}

\section{Model Recovery By Cold Boot Attack}

In this section, we present the practical cold boot attack for model recovery. We also present the threat model for the attack.

\subsection{Attack Target}
OpenVINO is a Python framework provided by Intel to interface with the NCS, load networks on it, run inference and get results \cite{openvino}. Before a model can be loaded on to the NCS, it has to first be transformed into OpenVINO's IR format.
Models in supported frameworks like TensorFlow, Keras \textit{etc}. can be converted into the IR format. Once converted, the IR format models are stored as two files: one containing the model architecture in XML format as a \texttt{.xml} file and the other containing the weights of the model as little-endian hexadecimal numbers in a \texttt{.bin} file. The weights are stored sequentially with the first weight being the weight value of the first neuron (or first filter in case of convolutional layer), followed by the rest of the neurons in that layer and the bias. This continues for all the layers sequentially.

As an edge accelerator, the NCS needs to be paired with a host Linux microprocessor. The baseline setup uses Raspberry Pi as the host computer. \emph{To load the model on to the NCS to perform inference, the IR files need to be first read and parsed by OpenVINO on the host device. Even if the \texttt{.xml} and \texttt{.bin} are encrypted when saving, they have to be decrypted before loading on to the RAM and parsed by OpenVINO \cite{openvino_security}. This means that the decrypted model weights and architecture can be retrieved from the contents of the host RAM by performing a cold boot attack.}

\begin{figure}[t]
\centering
\captionsetup{justification=centering}
\subfloat[NCS on Main Processor]
{
	\includegraphics[width=0.28\textwidth]{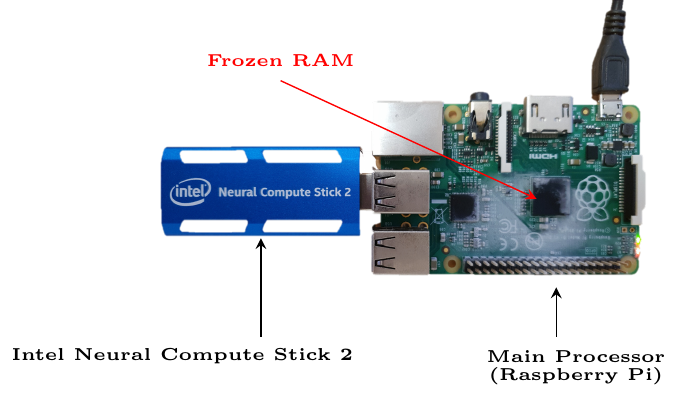}
	\label{fig:okdo_board}
}
\subfloat[Air Duster]
{
    \includegraphics[width=0.10\textwidth]{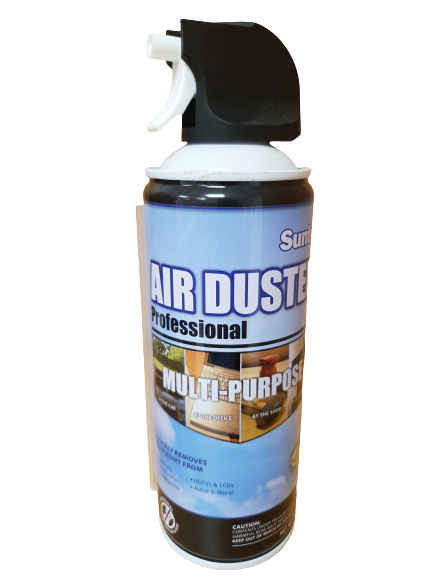}
    \label{fig:board_setup}
}
    \vspace{-2mm}
\caption{Cold Boot Attack against EdgeML on Raspberry Pi.}
\label{fig:cba_setup}
    \vspace{-3mm}
\end{figure}

\subsection{Threat Model}

The target victim is an edge device performing inference with trained model \textsf{M} which is of proprietary nature. The model inference is accelerated using NCS with Raspberry Pi as the host device. We expect that the designer has implemented standard security features, such as encryption of the SD card image, blocking unauthorised access to Raspberry Pi using password protection, OpenVINO model encryption \textit{etc}. %

The adversary aims at recovering a model \textsf{M'} comparable to model \textsf{M} of the victim. The first objective for the adversary is to achieve a \emph{task accuracy extraction} based on the previously stated taxonomy\cite{jagielski2020high} as it allows omitting payments towards pay per use licence. Possibility of a stronger extraction can always benefit the adversary and is explored later in the paper.
The adversary has physical access of the victim device to perform a cold boot attack in order to recover \textsf{M'}. The process of cold boot attack is explained in next section. Due to the nature of cold-boot attack, the recovered model will have some errors and thus, the model errors must be corrected to a desired threshold. Once the equivalent model \textsf{M'} is recovered, the victim device can be rebooted and restored to function normally. The model \textsf{M'} can be run on an independent device without incurring any licensing cost to the adversary. With regards to recovery of the parameter, we observed that in most trained models, the values (such as weights) typically fall in the range of \([-5, +5]\). This information could then be utilized by the adversary when performing model correction. This range can be suitably adapted if required.

\subsection{Practical Details on Cold Boot Attack}

As the OpenVINO framework runs on host computer, it handles the decryption of the model, conversion of model parameters to IR format and communicates with NCS. This requires the model to be stored in the SRAM of host computer in decrypted format. As stated earlier, the host computer readily used with NCS is Raspberry Pi. 

The aim of the cold boot attack is to recover the model from SRAM of the Raspberry Pi (Model B+). To perform the attack, we need to identify the boot sequence of Raspberry Pi.
Won \textit{et al.}~\cite{won2020practical} showed an attack on Raspberry Pi for recovery of memory content which was image data in their use case. We use a similar approach in the following.

The boot sequence starts with turning on the GPU and first stage bootloader is executed from the ROM. The first stage boot loader writes second stage boot loader to L2 cache and executes it to enable SRAM. The L2 cache is initialized by the bootloader preventing cold boot attack but this is not the case with SRAM which remains uninitialised. An adversary with physical access to device executing inference with the loaded model, freezes the RAM mounted on the host CPU. The RAM can be frozen using an air duster as shown in Fig.~\ref{fig:cba_setup}. The adversary then swaps the victim SD card with his own SD card to run a malicious image. The malicious image runs the boot sequence until second stage bootloader and dumps the uninitialised SRAM content. Once the RAM content is dumped, the device can be restored to allow normal operation with victim's SD card. \textit{The attack is extremely low cost.}

We perform the cold boot attack with OpenVINO framework and model inference on NCS. The target of the attack is to recover the decrypted model parameters which exist in the Raspberry Pi SRAM, while the inference is requested on the NCS. 

To measure the effectiveness of the attack, we compute the recovery ratio of the RAM data. This is measured in terms of decay models~\cite{halderman2009lest} where $\rho_{0}$ measures the probability of a original bit value 1 flipping to 0 and $\rho_{1}$ measures vice versa. We observe ($\rho_{0}$ and  $\rho_{1}$) as  ($0.0000027$ and  $0.00000009)$ respectively.
Compared to previous works like~\cite{halderman2009lest}, our model is recovered with a very low error rate. Previous reported numbers for $\rho_{0}$ and $\rho_{1}$ at 1\% and 0.1\% respectively or orders of magnitude higher than our case.

We further investigate the robustness of our proposed cold boot attack. The air duster used in our experiments can take down the temperature to levels as low as \SI{-30}{\celsius}. However, while operating under standard laboratory temperature (\SI{24}{\celsius}), the chip temperature starts to  rise up fast. %
In Fig.~\ref{fig:cba_result_temp}, the recovery ratio is plotted against temperature. At \SI{-30}{\celsius}, we could achieve a recovery ratio about 99.96\%. %
This degrades quickly to 99\% at \SI{-20}{\celsius} and 62\% at \SI{-10}{\celsius}.

Another interesting parameter for the proposed attack is the position of bit errors in the victim memory dump. If the error occur in random position, a simple repetition of the attack more than twice (when threat model allows) followed by a majority voting can reduce the error rates. To investigate this, we repeated the same attack 5 times under same conditions (\SI{-30}{\celsius}).
We quantify this in form of a $5\time 5$ cross-correlation similarity matrix for the 5 trials as shown in Fig.~\ref{fig:cc_result}.
A high correlation indicates that error often affect the same bit positions across multiple trials, thus preventing error correction by majority voting.
This could be linked to the memory chip hardware properties. The error rate with and without majority voting was reported at $\approx 0.04\%$, indicating that repetition does not help much.

\begin{figure}[t]
    \centering
    \captionsetup{justification=centering}
    \subfloat
    {
    	\includegraphics[width=0.23\textwidth]{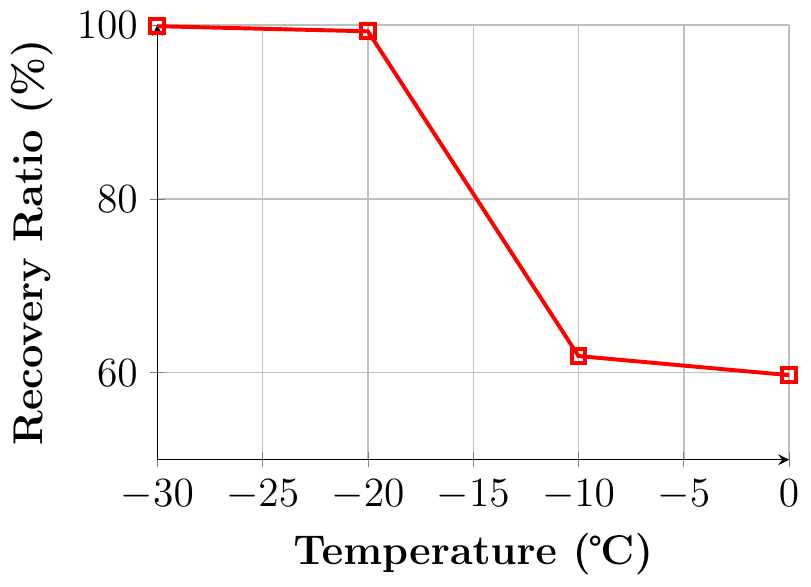}
    	\label{fig:cba_result_temp}
    }
    \subfloat
    {
    	\includegraphics[width=0.23\textwidth]{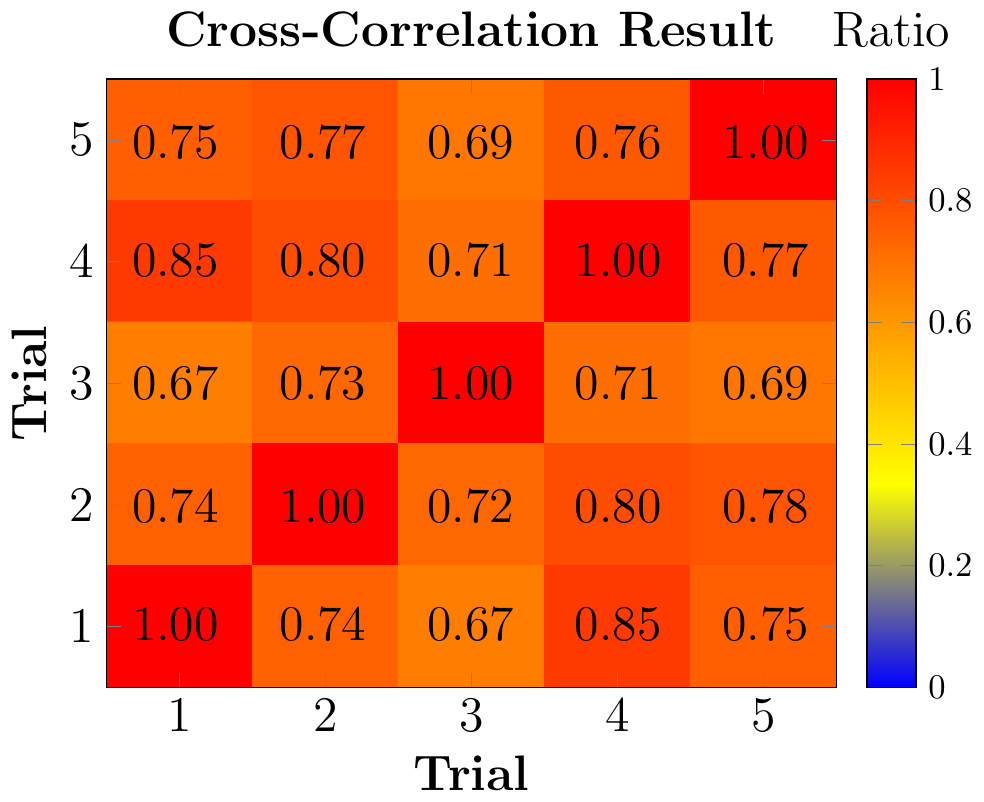}
    	\label{fig:cc_result}
    }
    \caption{Impact of temperature on the recovery Ratio for cold boot attack (left) and Cross-Correlation for 5 trials (right).}
    \label{fig:cba_new_result}
\end{figure}

\section{Baseline Model Recovery}
\label{sec:base}
Once the RAM dump is retrieved, the next step is to recover the model from the RAM dump. Due to errors in extraction, the recovered weights can be different from the weights of the original model.

\subsection{Recovering Model from RAM Dump (\texttt{.xml} + \texttt{.bin})}

The RAM dump as recovered with the cold boot attack must be interpreted to recover the target model \textsf{M'}. Due to the error introduced in RAM dump during the process of cold boot attack, the exact model \textsf{M} cannot be directly recovered. In the following, we explain the method to perform a first stage error correction to recover \textsf{M'} from the RAM dump.

\subsubsection{Architecture Recovery} The errors are distributed over the \texttt{.xml} file and the \texttt{.bin} file, which must be corrected. The \texttt{.xml} can be corrected easily due to its well-defined structure.  The \texttt{.xml} file or the architecture of the model starts and ends with a $<$\texttt{net}$>$ tag. Other tags include the $<$\texttt{layer}$>$ and $<$\texttt{edges}$>$ tags which contains information about each layer and how they are connected. There is also a $<$\texttt{cli}\_\texttt{parameters}$>$ tag that contains metadata about the model and framework specific information. Recovering the $\texttt{xml}$ can be done by searching for the starting and ending $<$\texttt{net}$>$ tags and extracting everything in between. The recovery can be verified by making sure that all opened tags were closed. This can be helpful in case a part of the architecture is present in a different part of the RAM dump. Fig.~\ref{fig:code} shows an example of recovered model architecture with errors highlighted in red. These errors can be fixed using the previous approach and implemented by a combination of \texttt{XML} syntax checker and a dictionary.

\begin{figure}[ht]
\scriptsize
 \begin{Verbatim}[commandchars=\\\{\}]
 <net name="simpl\red{a}_ffnn" \red{w}ers\red{)}on="7>
   <layer\red{S}>
     <layer id="0" name="sequ\red{a}ntia\red{h[}1_input" 
     type="I\red{.}put">
       <\red{m}utput>
         <port iD="0" precisio\red{N}="FP32">
           <dim>1\red{4}/di\red{M}>
           <dim>3</dim>
         <\red{.}port>
       </output>
     </layer\red{:}
   </layers>
 </net>
 \end{Verbatim}
\caption{Sample Erroneous XML Snippet}
\label{fig:code}
\end{figure}

\subsubsection{Weight Recovery} Next, the weights must also be recovered from the \texttt{.bin} file. The number of weight values can be calculated from the recovered architecture. As stated before, the weights are stored sequentially from input to output, neuron by neuron. To recover the weights from the RAM dump, we search for consecutive little-endian floating point values which are not valid UTF-8 characters in Alg.~\ref{alg:weight_recovery}. We perform the check for every two floating point values or 8 potential UTF-8 characters. If those 8 characters do not form a valid UTF-8 sequence, then we consider those two weight values as a part of the model and appended to our model. The search for weights is terminated when the number of recovered weights matches the expected number of weights from the architecture recovery. Some errors can render the weights out of range like infinity or NaN, leading for model inference to fail returning a NaN output. To counter such errors, the out of range weights must be corrected to bring them in acceptable ranges. The IEEE-754 floating point standard allocates the first bit for the sign, the next 8 bits for the exponent and the final 23 bits for the mantissa. Usually, a change in the mantissa of the weight will result in a small magnitude change and will be difficult to identify. Similarly, a change in the sign bit will also be difficult to spot. However, a single bit change in the exponent can result in a large magnitude change of the weight that can be easily identified. 

As described in threat model, we consider weights that are not in the range of -5 to +5 as incorrect \cite{DBLP:conf/uss/BatinaBJP19}. In addition to that, we also consider weights with an absolute value less than $10^{-5}$ to also be incorrect.
Once the erroneous weights are identified, large weight values are divided by $2$ until its value is within the range of $[-5, +5]$ and small weights are multiplied by $2$ until their absolute value is more than $10^{-5}$.

Note that, the recovered weights are still erroneous. For our experiments, the weights are affected by an error rate of $0.04\%$. In next subsection, we investigate the impact of this erroneous model recovery on model accuracy. %

\begin{algorithm}
\scriptsize
\SetAlgoLined
\textbf{Input:} RAM Dump(D), Total Weights(T)\;
\textbf{Initialization:} \textit{count = 0}, \textit{weight\_array=[]}\;
\While{bit in D}
{
    \While{count $\leq$ T}
    {
        bits = Read next 64 bits from D\;

        \eIf{not valid\_UTF8(\textit{bits})}
        {
            \textit{count}=\textit{count}+$2$\;
            \textit{weight\_array}.append(\textit{bits})\;
        }
        {
            \textit{count}=$0$\;
            \textit{weight\_array=[]}\;
        }
    }
    \textit{range\_count=}Count(\textit{weight\_array} in Range(-5, 5))\;
    \textit{range\_percent}=\textit{range\_count}/Length(\textit{range\_count})\;
    \eIf{range\_percent $\geq 0.9$}
    {
        \textit{break}\;
    }
    {
        \textit{count = 0}\;
        \textit{weight\_array=[]}\;
    }
}
\caption{Weight Recovery Procedure}
\label{alg:weight_recovery}
\end{algorithm}

\subsection{Target Models \& Attack Metrics}
We perform this attack on models trained on CIFAR10. We use the Base, Base (Wide), Base (Dropout), Base (PReLU), LeNet5 and LeNet5 (Dropout) architectures from \cite{hong2019terminal} and train them on CIFAR10. %
The error rate during our cold boot attack is \textbf{$\rho_{0}$} ($1 \rightarrow{} 0$) $= 0.00027\%$ and \textbf{$\rho_{1}$} ($0 \rightarrow{} 1$) $=0.000009\%$. We study the impact of such error rates on model accuracy. %

We use the Relative Accuracy Drop (RAD~\cite{hong2019terminal}) to quantify the drop in performance of the model:
\begin{equation}
    RAD = \frac{Acc_{M}-Acc_{M'}}{Acc_{M}},
\end{equation}
where \textsf{M} and \textsf{M'} are original and recovered models respectively.

\subsection{Experimental Results}

The baseline attack was performed 100 times for each model. The architecture and weights are recovered sequentially. First the architecture is recovered from the RAM dump followed by error correction as described earlier. The correction process is trivial and we were able to recover the architecture every time, leading to a \textbf{100\% success}.  The recovery of the architecture provides information on the number of weights. The weight recovery follows Algo~\ref{alg:weight_recovery} and contain errors at a rate of 0.04\%, which corrects the out of range errors previously mentioned. The errors are not further corrected for this experiment. 
Fig.~\ref{fig:cifar_rad_low} shows the result for the models trained on CIFAR10. 
It reports that the recovered model \textsf{M'} performs very closely to original model \textsf{M}, with a maximum RAD of 0.005. in some cases, the induced error results in a slight improvement of accuracy.

\begin{figure}[t] 
\centering
\includegraphics[width=0.8\linewidth]{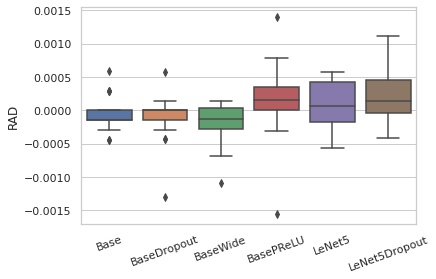}
\caption{RAD for recovered model with low error rates.}
\label{fig:cifar_rad_low}
\end{figure}

\section{Model Recovery With Higher Error Rates: Knowledge Distillation}

In the previous sections, we reported a successful cold boot base model recovery attack where the accuracy of recovered model \textsf{M'} was fairly close to \textsf{M}, leading to a task accurate extraction. The recovery was performed on our experimental test bed where we achieved extremely low error rates. However, if we compare to the previously reported cold boot attacks the error rates can be as high as 1\%~\cite{halderman2009lest}. This is also seen in our experiments when recovery is done at \SI{-20}{\celsius}. To assess the feasibility of cold boot attacks in a much general scenario, in this section, we work with higher error rates as reported in~\cite{halderman2009lest} \textit{i.e.} \textbf{$\rho_{0}$} ($1 \rightarrow{} 0$) $= 1\%$ and \textbf{$\rho_{1}$} ($0 \rightarrow{} 1$) $=0.1\%$. As shown later, higher error rates result in higher RAD. Thus we propose a methodology based on knowledge distillation to correct \textsf{M'}.

\begin{table*}[]
\caption{KD Based Model Recovery on Different Datasets From~\cite{hong2019terminal} with Higher Error Rate}
\label{tab:main_results}
\resizebox{\textwidth}{!}{%
\begin{tabular}{c|c|cccccccccccc}
\multicolumn{2}{c|}{\textbf{\backslashbox{Data}{Models}}} & \multicolumn{2}{c}{\textbf{Base}} & \multicolumn{2}{c}{\textbf{BaseWide}} & \multicolumn{2}{c}{\textbf{BaseDropout}} & \multicolumn{2}{c}{\textbf{BasePReLU}} & \multicolumn{2}{c}{\textbf{LeNet5}} & \multicolumn{2}{c}{\textbf{LeNet5Dropout}} \\ \hline
\textbf{Training} & \textbf{Recovery} & \textbf{D1} & \textbf{D2} & \textbf{D1} & \textbf{D2} & \textbf{D1} & \textbf{D2} & \textbf{D1} & \textbf{D2} & \textbf{D1} & \textbf{D2} & \textbf{D1} & \textbf{D2} \\ \hline
\multirow{3}{*}{MNIST} & MNIST & 0.002956 & 0.0009176 & 0.004581 & 0.00101 & 0.001424 & 0.004884 & 0.003776 & 0.004286 & 0.002442 & 0.002238 & 0.002953 & 0.003361 \\
 & \begin{tabular}[c]{@{}c@{}}EMNIST\\ (12k)\end{tabular} & 0.008666 & \textbf{0.004078} & 0.033594 & 0.01211 & 0.025132 & 0.008445 & 0.009899 & 0.006531 & 0.00661 & \textbf{0.006003} & 0.00682 & \textbf{0.007027} \\
 & \begin{tabular}[c]{@{}c@{}}EMNIST\\ (6k)\end{tabular} & 0.01009 & 0.007952 & 0.02046 & 0.026977 & 0.02319 & 0.01058 & 0.01765 & 0.02061 & 0.007834 & 0.006919 & 0.01028 & 0.009064 \\ \hline
\multirow{3}{*}{CIFAR10} & CIFAR10 & 0.081696 & 0.09497 & 0.13403 & 0.1091 & 0.09066 & 0.077327 & 0.06794 & 0.06726 & 0.05269 & 0.04236 & 0.05515 & 0.045448 \\
 & \begin{tabular}[c]{@{}c@{}}STL10\\ (10k)\end{tabular} & 0.056496 & \textbf{0.05283} & 0.099543 & 0.08153 & 0.08217 & 0.055503 & 0.055001 & 0.05244 & 0.05864 & \textbf{0.03642} & 0.04907 & \textbf{0.030558} \\
 & \begin{tabular}[c]{@{}c@{}}STL10\\ (5k)\end{tabular} & 0.07600 & 0.076818 & 0.135556 & 0.136063 & 0.08810 & 0.09214 & 0.0796710 & 0.07899 & 0.04701 & 0.04533 & 0.03224 & 0.03703
\end{tabular}%
}

\end{table*}

The baseline model recovery method can help to identify and fix bit flips in the exponent of the weights. However, as there are 23 mantissa bits, they are more prone to errors. %
Changes in the mantissa results in a very small change in the magnitude of the weight. This makes it hard to identify and correct those weights using the baseline approach. Further, even though the changes are small, as shown in Fig.~\ref{fig:cifar_rad}, RAD can be as high as 0.7 (compared to 0.005 with low error rates).

\begin{figure}[t] 
\centering
\includegraphics[width=0.7\linewidth]{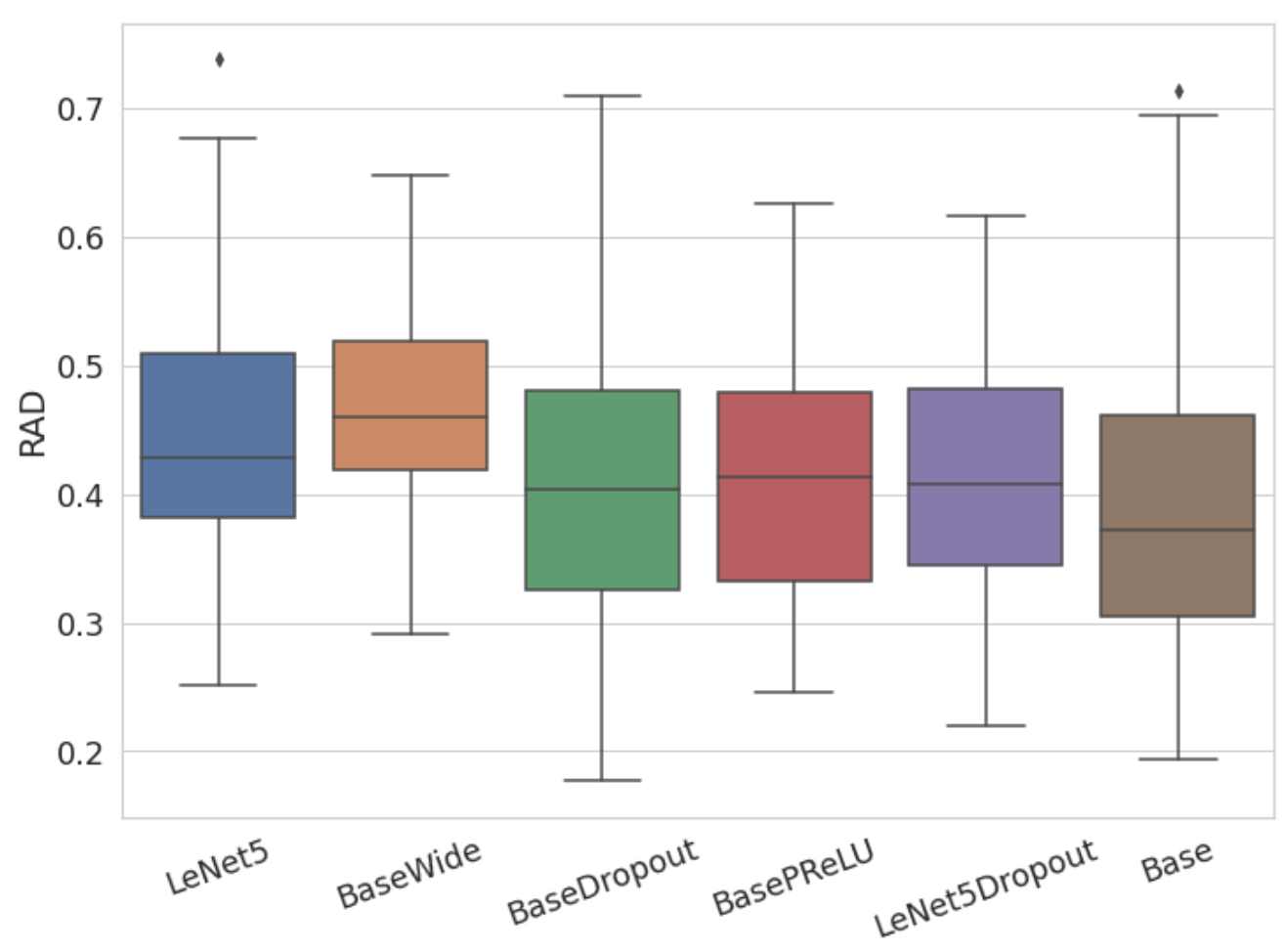}
\caption{RAD for recovered model with higher error rates.}
\label{fig:cifar_rad}
\end{figure}

We investigate several scenarios. The first scenario is where the original dataset is available to the attacker. Next, we also investigate a setting when original dataset is secret and  not available but a dataset similar to what the teacher model would have been trained with is used for training. Furthermore, to limit compute power of the attacker, we limit our training samples to only 10\% of the total samples in the dataset.

To improve the model accuracy, we adopt a two-step training approach to recover the accuracy of the student model. First, weights outside the range $[-5, +5]$ and weights with absolute value less than $10^{-5}$ are converted to zero. This initial error correction was chosen since it was simpler and in our experiments it lead to comparable results as the previous method (division of weights until in range). Following that, the model is trained using two different KD techniques.

\subsubsection{Traditional Knowledge Distillation} Traditionally, KD uses softmax temperature and intermediate outputs to train models. The extra information helps the student better learn from the teacher and improves student performance. However, in our application, since intermediate outputs and non-softmax outputs are not available to the attacker, these methods are not used. Instead, only the softmax outputs from the teacher model is used to train the student model, referred to as \textbf{D1}.

\subsubsection{Dropout Knowledge Distillation} In addition to improve accuracy, intermediate outputs and softmax temperature act as regularizers and prevents the student model from overfitting \cite{gou2021knowledge}. However, due to our constrained training setting, these techniques cannot be used. Additionally, since we use a smaller sample of recovery data different from the original training set, regularization is needed to prevent the model from overfitting. To counter these problems, we introduce Dropout KD as a regularization method, where random gradients of each layer are dropped out during the training of the student model (referred as \textbf{D2}). Weights whose gradients have been dropped are not updated.

KL-Divergence loss between the softmax outputs of the teacher and student model is used as the loss function for both D1 and D2.
For models trained on MNIST, we use the letters subset from the EMNIST dataset as the recovery dataset. For CIFAR10 and ImageNet models, the unlabelled subset from the STL10 dataset is used.

To recover the accuracy of the original MNIST models, training was done using Adam optimizer with a batch size of 32 and a learning rate of $10^{-3}$ for 30 epochs. The learning rate is halved every 10 epochs. For the CIFAR10 trained models, Adam with a batch size of 128 and a learning rate of $10^{-4}$ is used. The learning rate is reduced by a factor of $0.9$ every 10 epochs and the models are trained for 50 epochs.

\subsubsection{Experimental Results}
We report the model recovery results for all the models in Tab.~\ref{tab:main_results}. We first report the RAD when the attacker has access to the original training dataset. In this case, only $10\%$ of the original training data is used for recovery. In addition to that, we also perform recovery using EMNIST (6k) and STL10 (5k) having same dataset size as the original dataset. Here 6k and 5k refer to randomly chosen 6000 and 5000 samples in training dataset. Next, we also investigate the scenario when attacker has access to larger dataset which is different from original dataset. In this case, we use $10\%$ EMNIST(12k) and STL10(10k). The RAD is calculated on the test set of the teacher model.

Not to surprise, access to original dataset, gives the best results. Fortunately, using more training samples also results in a decrease in the RAD irrespective of the recovery set used pointing to practical possibility of model recovery. This can be seen in Fig. \ref{fig:lenet_data_percentage} where \textbf{D1} was used as the training approach.

Further, using \textbf{D2} with a dropout value of $0.5$, in general, results in lower RAD especially when using more data to train. In addition, we also see that our method is not affected by the use of Dropout or PReLU during training and the depth of the network does not impact accuracy recovery.

On MNIST, the best result is on the Base model using $10\%$ of unlabelled EMNIST data as the recovery set and \textbf{D2} as the training paradigm. The LeNet5 and LeNet5 Dropout models have the second and third best recovery results using the same data and training paradigm.

On the other hand, on CIFAR10, the best RAD are on the LeNet5 and LeNet5 Dropout models using STL10 with 10k samples as the recovery set and \textbf{D2} as the training method.

\begin{table}[t]
\caption{Comparison of Knowledge Distillation (D2) vs. Re-Training. Results are in the form of Epochs/RAD}
\label{tab:kd_vs_training}
\resizebox{0.5\textwidth}{!}{
\begin{tabular}{c|ccc|c|c|c|c}
 & \multicolumn{5}{c|}{\textbf{Knowledge Distillation Training}} & \multicolumn{2}{c}{\textbf{Training from Scratch}} \\ \hline
\textbf{Recovery Data} & \multicolumn{3}{c|}{\textbf{STL10}} & \multicolumn{2}{c|}{\textbf{CIFAR10}} & \multicolumn{2}{c}{\textbf{CIFAR10}} \\ \hline
\textbf{Data Split} & \multicolumn{1}{c|}{0.1} & \multicolumn{1}{c|}{0.5} & 1 & 0.1 & 0.5 & 0.1 & 0.5 \\ \hline
\textbf{Base} & \multicolumn{1}{c|}{10/0.04} & \multicolumn{1}{c|}{13/0.01} & 6/0.003 & 10/0.036 & 7/0.017 & 14/0.34 & 7/0.29 \\ \hline
\textbf{BaseWide} & \multicolumn{1}{c|}{14/0.09} & \multicolumn{1}{c|}{7/0.03} & 3/0.01 & 12/0.05 & 15/0.018 & 16/0.34 & 14/0.16 \\ \hline
\textbf{BaseDropout} & \multicolumn{1}{c|}{20/0.047} & \multicolumn{1}{c|}{11/0.01} & 4/0.01 & 9/0.07 & 8/0.025 & 11/0.36 & 9/0.22 \\ \hline
\textbf{BasePReLU} & \multicolumn{1}{c|}{11/0.03} & \multicolumn{1}{c|}{7/0.008} & 2/0.003 & 8/0.04 & 6/0.006 & 12/45.11 & 26/0.04 \\ \hline
\textbf{LeNet5} & \multicolumn{1}{c|}{16/0.02} & \multicolumn{1}{c|}{10/0.01} & 4/0.01 & 9/0.028 & 5/0.0018 & 40/29.32 & 17/48.27 \\ \hline
\textbf{LeNet5Dropout} & \multicolumn{1}{c|}{9/0.01} & \multicolumn{1}{c|}{7/0.001} & 2/0.001 & 4/0.034 & 3/0.014 & 53/27.07 & 11/44.24
\end{tabular}
}
\end{table}

\subsubsection{Comparison With  Re-training}
To justify the use of distillation, it is compared with retraining the model from scratch using the same dataset. We report with CIFAR 10 only being harder than MNIST. As shown earlier, we can recover the architecture with 100\% success, which is now retrained from scratch as recovered weights are ignored. The results comparing the proposed distillation method (\textbf{D2}) with retraining (with successful architecture recovery) at different dataset splits can be seen in Tab. \ref{tab:kd_vs_training}. 

When training the same architecture from scratch with a smaller percentage of the training data, the accuracy recovery is worse overall as compared to training with \textbf{D2} as can be seen from the higher values of RAD. The epochs taken for the model to converge (loss not decreasing for more than 2 epochs) when training from scratch is also more in general compared to KD. When training with KD, a higher percentage of either the same training data or a similar unlabelled training data (STL10) results in a better RAD as well as fewer epochs to converge in general, with models trained on same data getting better results. Thus, KD provides advantage over recovering just the architecture and retraining.

\begin{figure}[t] 
\centering
\includegraphics[width=0.7\linewidth]{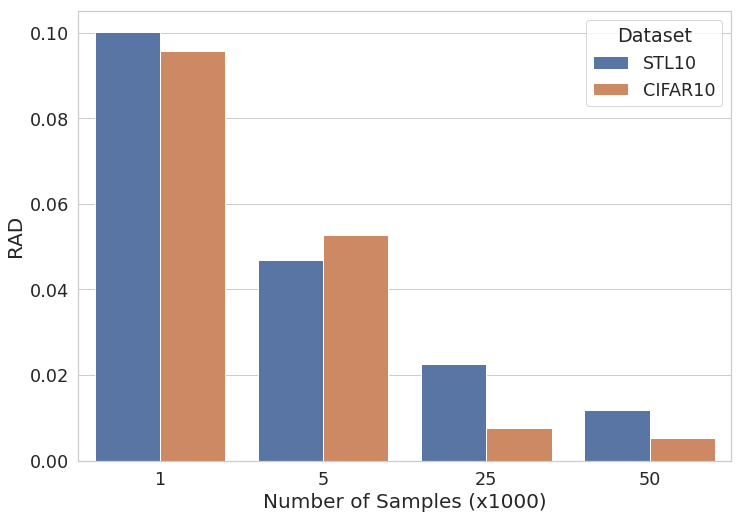}
\caption{RAD for CIFAR10 LeNet5 for Different Percentages of Training Dataset using D1 Training Paradigm}
\label{fig:lenet_data_percentage}
\end{figure}

\begin{table}[b]
\caption{Model Recovery on Pretrained Models}
\label{tab:transfer_results}
\resizebox{\linewidth}{!}{
\begin{tabular}{c|c|c|c|c|c|cc}
\multicolumn{2}{c|}{\textbf{\backslashbox{Data}{Models}}} & \multicolumn{2}{c|}{\textbf{AlexNet}} & \multicolumn{2}{c|}{\textbf{VGG16}} & \multicolumn{2}{c}{\textbf{ResNet18}} \\ \hline
\textbf{Training} & \textbf{Recovery} & \textbf{D1} & \textbf{D2} & \textbf{D1} & \textbf{D2} & \multicolumn{1}{c|}{\textbf{D1}} & \textbf{D2} \\ \hline
\multirow{2}{*}{\begin{tabular}[c]{@{}c@{}}Tiny\\ ImageNet\end{tabular}} & \begin{tabular}[c]{@{}c@{}}Tiny\\ ImageNet\end{tabular} & 0.08246 & 0.08366 & 0.08532 & 0.09834 & \multicolumn{1}{c|}{0.0289} & 0.09163 \\ \cline{2-8} 
 & \begin{tabular}[c]{@{}c@{}}STL10\\ (unlabelled)\end{tabular} & 0.08488 & 0.08717 & 0.09427 & 0.09892 & \multicolumn{1}{c|}{0.0935} & 0.09220
\end{tabular}
}
\end{table}

\subsubsection{On Transfer Learning}
Most deployed models do not have a custom architecture. Instead transfer learning on popular pre-trained ImageNet models is done to increase the accuracy on a separate dataset. In this case, the adversary has to only recover the fully connected (FC) layers.

We train three models: AlexNet, VGG16 and ResNet18 on Tiny ImageNet and try to recover the original accuracy using $50\%$ (or $50$k images) of the training set from Tiny ImageNet as well as the unlabelled set from STL10. The results are in Tab.~\ref{tab:transfer_results}. Here, the accuracy recovery works better when the same dataset that was used for training the model is used during recovery. However, the magnitude of RAD is comparable to when training the whole model. This could be attributed to fully connected layers where weights are recovered with higher error (as shown in next section).

\section{Further Discussions}

In this section, we explore the extent of our extracted model against various advanced attacks. 

\subsubsection{Exact Extraction}
The adversary model presented before aims at task accuracy extraction. However, for some applications exact extraction may be desirable. The use of KD in model recovery causes the weights in \textsf{M'} to take a different distribution as compared to \textsf{M}.
We conducted experiments and observed that it is possible to recover the original weights by training the models slower, with a smaller learning rate. To see how similar the weights of a layer in \textsf{M'} are as compared to the weights in \textsf{M}, we calculate the Relative Layer Norm defined as:
\begin{equation}
    Layer Norm = \frac{Norm(M - M')}{Norm(M)},
\end{equation}
where a lower Layer Norm means that weights are similar. When low error rates are encountered, the norm stays close to 1 in general (see Fig. \ref{fig:layer_norm_low}), which indicates that the distribution has minor impact.

For higher error rates, we perform the layer activation recovery by training with an Adam optimizer for 500 epochs with a learning rate of $10^{-5}$. We reduce the learning rate by $0.5\times$ every $50$ epochs. The recovery was done with the full unlabelled subset of the STL10 dataset on 4 models trained on CIFAR10 at high error rates. The results are in Fig. \ref{fig:layer_norm_high}.
In general the weights in the first and last layers can be recovered better than the weights in the middle layers. In addition, layers with more parameters like the FC layer are harder to recover and in some cases, even with slower training, the layer norm becomes greater than 1 meaning that the weights take on a different distribution. The higher error rates in FC layer also confirms the poor RAD of transfer learning.

\begin{figure}[t]
    \centering
    \captionsetup{justification=centering}
    \subfloat[][With low error rates]
    {
    	\includegraphics[width=0.5\linewidth]{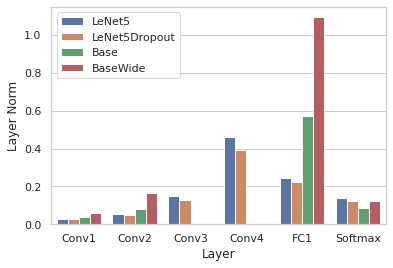}
        \label{fig:layer_norm_low}
    }
    \subfloat[][With higher error rates]
    {
    	\includegraphics[width=0.5\linewidth]{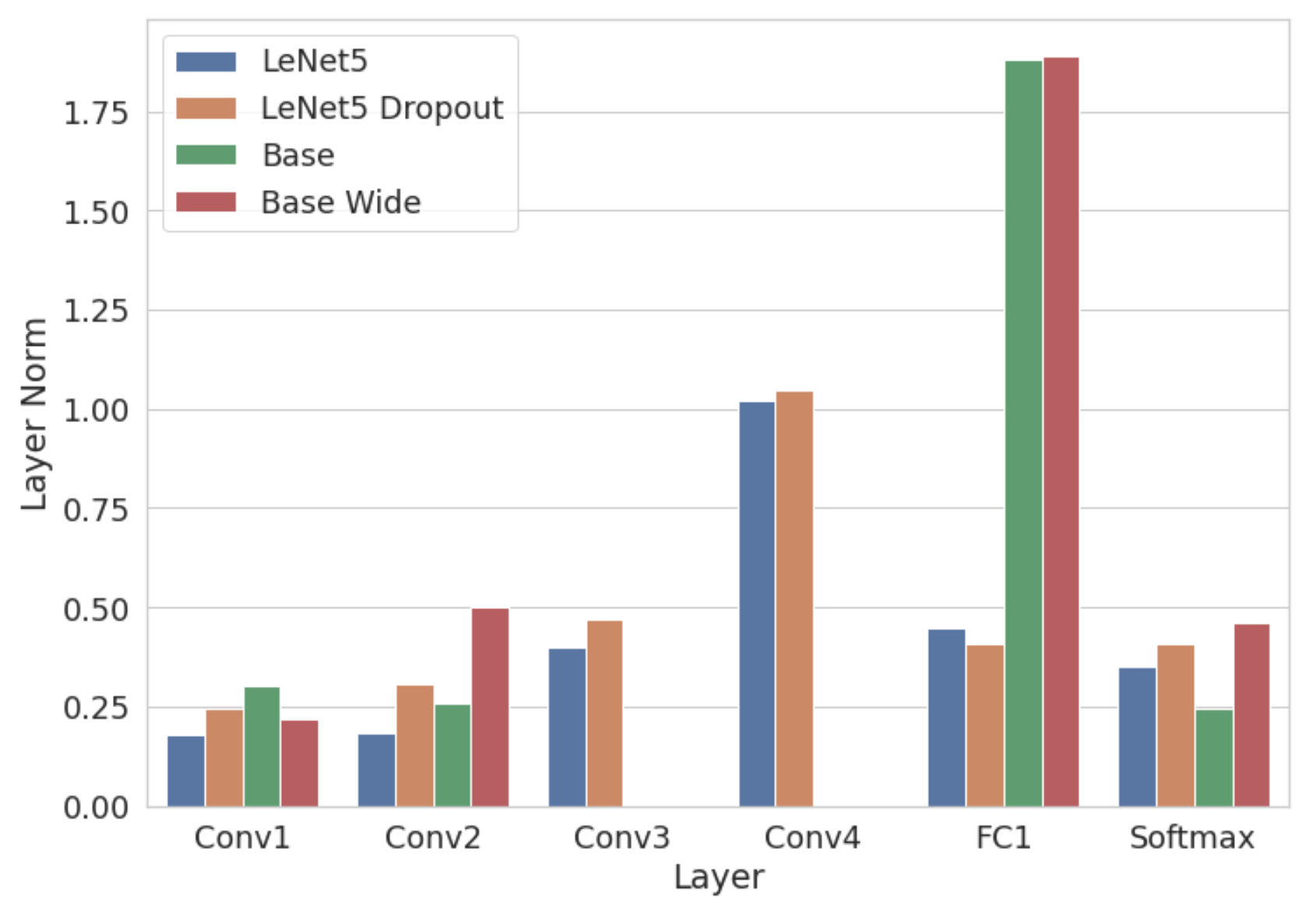}
        \label{fig:layer_norm_high}
    }
    \caption{Layer Norm After Training to Recover Original Weights of CIFAR10 Models}
    \label{fig:layer_norm}
\end{figure}

\subsubsection{High Fidelity Extraction}
High-fidelity extraction is weaker than exact extraction but still enables advanced adversary capabilities. %
We follow the approach of~\cite{jagielski2020high} and test the fidelity of the retrieved model by assessing the transferability of adversarial examples. High-fidelity models ensure maximal transfer of adversarial examples.  
Here, we use the Fast Gradient Signed Method (FGSM) attack~\cite{DBLP:journals/corr/GoodfellowSS14}, to investigate the transferability of adversarial example on the recovered model. We investigate LeNet-5 model trained on CIFAR-10 dataset. For low error rates, the recovered model achieved averaged RAD of 0.002. The average fidelity, when computed over multiple runs, is around 97.66\% on adversarial examples, with $\epsilon = 0.01$ and $0.1$, where $\epsilon$ denotes degree of perturbations. in other words, 97.66\% of the adversarial examples transfer from \textsf{M} to \textsf{M'}. For high-error rates (followed by model recovery by distillation), the average fidelity is around 81.78\% and 89.20\% on adversarial examples, with $\epsilon = 0.01$ and $0.1$ respectively. This matches previous results where weights after distillation deviate from the original model, thus degrading fidelity.

\subsubsection{Application to Model Inversion}

Model inversion (MI) is a technique which enables an adversary to recover the training data from model parameters. %
We investigated the potential of MI on \textsf{M'} following the approach of Maximum a Posterior (MAP) principle was suggested in~\cite{fredrikson2014privacy}.
We generated training samples from \textsf{M'} using the methodology and code proposed by~\cite{fredrikson2015model} and error rates as used in Sec.~\ref{sec:base}.
The generated samples are then classified using the original model \textsf{M} to measure the accuracy.
While \textsf{M} could achieve an accuracy of 0.91 originally, the samples generated from \textsf{M'} resulted in lower accuracy, leading to a RAD of 0.47 on average. 
We observed the worst case and best case accuracy to be 0 and 0.87 respectively, which depends on the position of the errors. As the RAD with low error rates is already high, we do not repeat it for higher error rate. In conclusion, model inversion is difficult on erroneous model, even when error rates are low. %

\subsubsection{Mitigation}

To mitigate this attack two approaches can be adopted. From the hardware side, the boot sequence can be fixed to initialize the SRAM at each boot-up. However, an advanced adversary can disable such initialization by semi-invasive means. Alternatively, secure hardware enclaves can be adopted for sensitive model execution, which may result in some performance overheads. At the software level, computation over encrypted model can be adopted to counter such attacks, at the cost of performance.

\section{Conclusions}
We demonstrate  practical cold boot attack and model recovery on a commercial EdgeML device setup with minimal accuracy loss and high fidelity. Considering higher error rates in the previous papers, we propose a methodology for model correction based on KD. Moreover, with dropout KD, the model can be recovered with comparable accuracy, even without access to original training datasets, by using similar unlabeled dataset. We test our method on 6 different architectures from \cite{hong2019terminal} as well as on 3 pre-trained models trained on MNIST, CIFAR10 and Tiny ImageNet. Future work will focus on scalability to larger networks and other network parameters like pruning, choice of activation function. %

\section*{Acknowledgment}
This research is supported by the National Research Foundation, Singapore, under its National Cybersecurity Research $\&$ Development Programme / Cyber-Hardware Forensic $\&$ Assurance Evaluation R$\&$D Programme (Award: NRF2018NCR-NCR009-0001).

\bibliographystyle{IEEEtran}
\bibliography{IEEEfull.bib}

\end{document}